\documentclass[modern]{aastex631}

\begin{document}

\title{
Crowded Field Photometry with Rubin: Exploring 47 Tucanae with Data Preview 1
}


\author[0000-0001-6320-2230]{Tobin M. Wainer}
\affiliation{Department of Astronomy and the DiRAC Institute, University of Washington, 3910 15th Avenue NE, Seattle, WA 98195, USA}

\author[0000-0002-0637-835X]{James R. A. Davenport}
\affiliation{Department of Astronomy and the DiRAC Institute, University of Washington, 3910 15th Avenue NE, Seattle, WA 98195, USA}

\author[0000-0001-8018-5348]{Eric C. Bellm}
\affiliation{Department of Astronomy and the DiRAC Institute, University of Washington, 3910 15th Avenue NE, Seattle, WA 98195, USA}

\author[0000-0001-5538-0395]{Yuankun (David) Wang}
\affiliation{Department of Astronomy and the DiRAC Institute, University of Washington, 3910 15th Avenue NE, Seattle, WA 98195, USA}

\author[0000-0003-3287-5250]{Neven Caplar}
\affiliation{Department of Astronomy and the DiRAC Institute, University of Washington, 3910 15th Avenue NE, Seattle, WA 98195, USA}

\author[0009-0006-9972-6937]{Elliott S. Burdett}
\affiliation{Department of Astronomy and the DiRAC Institute, University of Washington, 3910 15th Avenue NE, Seattle, WA 98195, USA}

\author[0000-0003-2497-091X]{Nora Shipp}
\affiliation{Department of Astronomy and the DiRAC Institute, University of Washington, 3910 15th Avenue NE, Seattle, WA 98195, USA}

\author[0009-0001-9549-0457]{John K. Parejko}
\affiliation{Department of Astronomy and the DiRAC Institute, University of Washington, 3910 15th Avenue NE, Seattle, WA 98195, USA}

\author[0009-0009-0252-5808]{Gray Thoron}
\affiliation{Department of Astronomy and the DiRAC Institute, University of Washington, 3910 15th Avenue NE, Seattle, WA 98195, USA}

\author[0009-0006-8050-4795]{Eric Butler}
\affiliation{Department of Astronomy, Columbia University, 538 W. 120th Street, Pupin Hall, New York, NY 10027, USA}

\author[0009-0005-6383-8892]{Maya Salwa}
\affiliation{Department of Astronomy and the DiRAC Institute, University of Washington, 3910 15th Avenue NE, Seattle, WA 98195, USA}

\author[0000-0002-0716-947X]{Erin Leigh Howard}
\affiliation{Department of Astronomy and the DiRAC Institute, University of Washington, 3910 15th Avenue NE, Seattle, WA 98195, USA}

\author[0000-0002-5012-3549]{Brianna Marie Smart}
\affiliation{Department of Astronomy and the DiRAC Institute, University of Washington, 3910 15th Avenue NE, Seattle, WA 98195, USA}

\author[0009-0003-1791-8707]{Wilson Beebe}
\affiliation{Department of Astronomy and the DiRAC Institute, University of Washington, 3910 15th Avenue NE, Seattle, WA 98195, USA}

\author[0000-0002-6823-7798]{Ishan F. Ghosh-Coutinho}
\affiliation{Department of Astronomy and the DiRAC Institute, University of Washington, 3910 15th Avenue NE, Seattle, WA 98195, USA}

\author[0000-0001-8418-3083]{Bob Abel}
\affiliation{Department of Astronomy and the DiRAC Institute, University of Washington, 3910 15th Avenue NE, Seattle, WA 98195, USA}

\author[0000-0001-5250-2633]{\v{Z}eljko Ivezi\'{c}}
\affiliation{Department of Astronomy and the DiRAC Institute, University of Washington, 3910 15th Avenue NE, Seattle, WA 98195, USA}

\begin{abstract}
We analyze imaging from Data Preview 1 of the Vera C. Rubin Observatory to explore the performance of early LSST pipelines in the 47 Tucanae field. 
The coadd-\texttt{object} catalog demonstrates the depth and precision possible with Rubin, recovering well-defined color magnitude diagrams for 47 Tuc Small Magellanic Cloud.
Unfortunately, the existing pipelines fail to recover sources within $\sim$28 pc of the cluster center, due to the extreme source density. 
Using Rubin's forced photometry on stars identified via Difference Imaging, we can recover sources down to $\sim$14 pc from the cluster center, and find 14744 potential cluster members with this extended dataset. While this forced photometry has significant systematics, our analysis showcases the potential for detailed structural studies of crowded fields with the Rubin Observatory.

\end{abstract}

\section{DP1 Analysis} 

Globular clusters (GCs) are critical 
testbeds for models of stellar evolution and Galactic formation history \citep[e.g.,][]{forbes_globular_2018}. As one of the most massive and nearby GCs, 47 Tucanae is a unique laboratory for such studies
\citep{baumgardt_accurate_2021}.
Using the Vera C. Rubin Observatory's Data Preview 1 \citep[DP1;][]{rubin_observatory_dp1_2025}, \citet{choi_47_2025} explored 47 Tuc against a background of both Milky Way and Small Magellanic Cloud (SMC) stars. We build on this work by analyzing where the coadd data is limited due to crowding, and how one could probe deeper into the cluster.

Here we compare two catalogs available through the Rubin Science Platform \citep{rospd_lsst_2025}: 1) the coadd-based \texttt{object} catalog, and 2) the \texttt{diaForcedSource} catalog, which provides forced photometry for every Rubin visit. 

\begin{figure}
    \centering
    \includegraphics[width=0.99\textwidth]{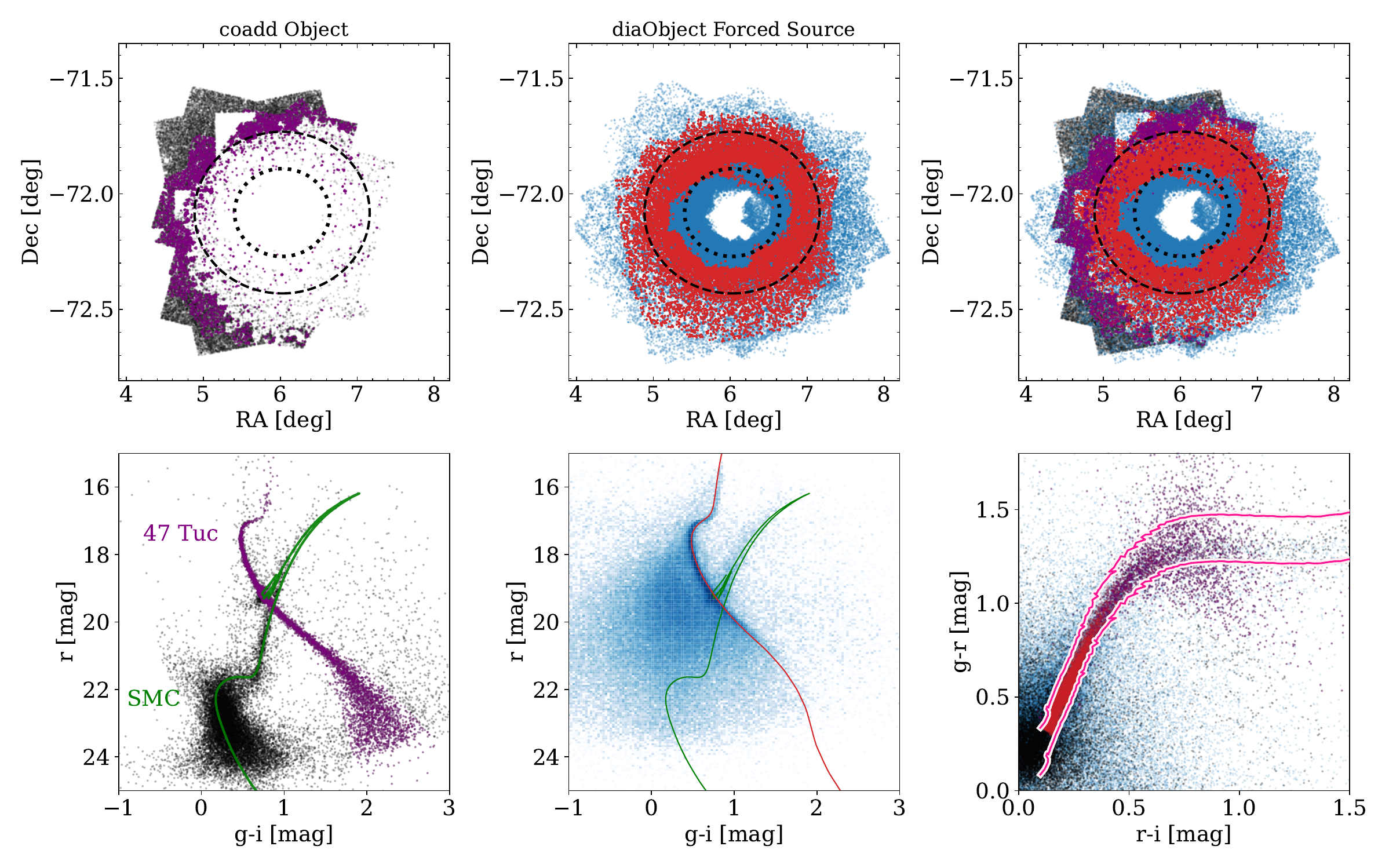}
    \caption{
    Top: Spatial distributions for 47 Tuc from the DP1 coadd-\texttt{object} catalog (left), and the  \texttt{diaForcedSource} median catalog (middle), with both shown for direct comparison (right).
    Approximate inner completeness radii of $\sim$28 and $\sim$14 pc (dashed and dotted lines, respectively) are shown.
    Bottom: CMDs (left, center) for the coadd-\texttt{object} (left) and \texttt{diaForcedSource} catalogs (middle). Color–color diagram (right) for both samples, with the stellar locus cut described in the text (pink lines).
    }
    \label{fig:cmd_spatial}
\end{figure}

\section{Coadd Objects}
Using the LSDB catalog framework \citep{caplar_using_2025}, we obtained measurements from the deep coadd-\texttt{object} catalog for 32011 objects within 5 degrees of 47 Tuc, with $25 > g > 14$ mag, and \texttt{Extendedness == 0}. 
We select possible cluster members using a combination of color-magnitude diagrams (CMD) and color-color cuts. We selected objects within a region centered around a MIST isochrone \citep{choi_mesa_2016, dotter_mesa_2016}, with $\log_{10}(\mathrm{age/yr}) = 10.3$, [Fe/H]=-0.78, and $m-M$=13.0. We hand-tuned the width of this region by the mean $r$-band uncertainty. 
To eliminate contamination, we required stars to be $<0.2$ mag from the known stellar color-color locus \citep{davenport_sdss-2mass-wise_2014}.
In total, we recovered 4506 candidate cluster members, comparable to the sample from \citet{choi_47_2025}.

The effects of crowding are unambiguous, with a sharp falloff in sources inside $\sim28$ pc of the cluster center.
However, the photometry outside this radius is excellent. The CMD shows a remarkably clean main sequence down to $r \sim 24$ magnitude, a well-defined turnoff, and is clearly separated from the SMC. The Rubin pipeline is capable of producing deep, science-grade photometry in moderately crowded fields, but cannot probe into the crowded cluster regime that makes clusters like 47 Tuc so interesting, where multiple populations, mass segregation, and dynamical evolution leave their imprint. To try and better probe this region, we explore another Rubin photometry approach.

\section{Forced Photometry}

To push deeper into the cluster, we use the \texttt{diaForcedSource} catalog, which provides forced PSF photometry on both the direct and difference images at the position of each object identified by the image differencing pipeline (DIAObject).
Many stars are detected by image differencing even in dense regions of 47 Tuc. Using forced measurements at these positions bypasses detection and deblending stages in the coadd-based pipeline, allowing us to probe a subset of objects  closer to the core. For each DIAObject, we compute the median of this photometry in each band across all available visits, resulting in a total of 147882 unique objects.

To ensure high quality detections, we required the median $gri$ photometric errors for each object be $<0.25$ mag. We further removed sources with only 1 detection any of in the $gri$ bands, and required the standard deviation of detections in each band to be $<0.25$ mag. While this may remove a small number of intrinsically variable sources, it eliminated objects with large outliers. In total we analyzed 116963 forced photometry objects. This sample detects stars much further into the cluster than the coadd-\texttt{object} data, but exhibits significant systematics as demonstrated in the CMD. A prominent plume of spurious sources appears blue of $(g - i) \sim 0.6$, which are inconsistent with any stellar population in the field (e.g. SMC or Milky Way), and spatially align with the region closest to the cluster core. 
These biases are not surprising in such crowded fields, as the forced PSF measurements are not deblended, and bad image subtractions can lead to centroid offsets of the DIAObjects.

Despite these limitations, the resulting CMD shows remarkably rich structure. The cluster isochrone is well sampled down to $r\approx21$ mag, and we recover sources nearly a factor of two deeper into the cluster ($\sim14$ pc) than the coadd-\texttt{object} catalog. The SMC is less prominent, but the giant and horizontal branches are still clearly visible in the CMD.

We applied the same CMD and color-color selection as before, and recovered 14744 possible members of 47 Tuc. The color-color cut is particularly important for eliminating contamination from blended photometry. Despite the shallower depth of these per-visit detections, this sample is more than 3x larger than detected in the coadd-\texttt{object} data. 

While the study of dense regions is limited by current Rubin photometric pipelines, we acknowledge the preliminary nature of DP1. All images were acquired during the brief LSST Commissioning-Camera observing run, and
47 Tuc was specifically chosen to test the performance of general-purpose Rubin pipelines in the presence of extreme crowding.
LSST will deliver more visits to crowded fields, enabling us to better probe the inner regions of such clusters. Our analysis serves not as a critique, but as a demonstration of the impact that photometric methodology can provide in crowded regimes. The contrast between the coadd-\texttt{object} and \texttt{diaForcedSource} catalogs underscores the degree to which algorithmic choices can affect completeness and reliability, especially in scientifically rich, but observationally extreme environments.

\vspace{1.25cm}
\begin{acknowledgments}
We acknowledge support from the DiRAC Institute in the Department of Astronomy at the University of Washington, supported through generous gifts from the Charles and Lisa Simonyi Fund for Arts and Sciences, Janet and Lloyd Frink, and the Washington Research Foundation.

This material is based upon work supported in part by the National Science Foundation through Cooperative Agreements AST-1258333 and AST-2241526 and Cooperative Support Agreements AST-1202910 and 2211468 managed by the Association of Universities for Research in Astronomy (AURA), and the Department of Energy under Contract No. DE-AC02-76SF00515 with the SLAC National Accelerator Laboratory managed by Stanford University. Additional Rubin Observatory funding comes from private donations, grants to universities, and in-kind support from LSST-DA Institutional Members.
\end{acknowledgments}

\bibliography{rubin_bib, Tobins_references}.
\bibliographystyle{mnras}

\end{document}